\documentclass[superscriptaddress, letterpaper,twocolumn,prl,showpacs,reprint]{revtex4}

\usepackage{graphicx,epsfig}
\usepackage{dcolumn}
\usepackage{bm}
\usepackage{textcomp}

\begin{document}

\title{Quasiparticle Mass Enhancement and Temperature Dependence of the Electronic Structure of Ferromagnetic SrRuO$_{3}$ Thin Films}

\author{D.E. Shai}
 \affiliation{Laboratory of Atomic and Solid State Physics, Department of Physics, Cornell University, Ithaca, New York 14853, USA}

\author{C. Adamo}
 \affiliation{Department of Materials Science and Engineering, Cornell University, Ithaca, New York 14853, USA}

\author{D.W. Shen}
\affiliation{Laboratory of Atomic and Solid State Physics, Department of Physics, Cornell University, Ithaca, New York 14853, USA}
 \affiliation{Department of Materials Science and Engineering, Cornell University, Ithaca, New York 14853, USA}
 \affiliation{Superconductor Applications State Key Laboratory of Functional Materials for Informatics, Chinese Academy of Sciences, Shanghai 200050, China}

\author{C.M. Brooks}
 \affiliation{Department of Materials Science and Engineering, Cornell University, Ithaca, New York 14853, USA}
 \affiliation{Department of Materials Science and Engineering, The Pennsylvania State University, University Park, Pennsylvania 16802, USA}
 
\author{J.W. Harter}
\affiliation{Laboratory of Atomic and Solid State Physics, Department of Physics, Cornell University, Ithaca, New York 14853, USA}

\author{E.J. Monkman}
\affiliation{Laboratory of Atomic and Solid State Physics, Department of Physics, Cornell University, Ithaca, New York 14853, USA}
 
\author{B. Burganov}
\affiliation{Laboratory of Atomic and Solid State Physics, Department of Physics, Cornell University, Ithaca, New York 14853, USA}
 
\author{D.G. Schlom}
 \affiliation{Department of Materials Science and Engineering, Cornell University, Ithaca, New York 14853, USA}
 \affiliation{Kavli Institute at Cornell for Nanoscale Science, Ithaca, New York 14853, USA}

\author{K.M. Shen}
 \email[Author to whom correspondence should be addressed: ]{kmshen@cornell.edu}
\affiliation{Laboratory of Atomic and Solid State Physics, Department of Physics, Cornell University, Ithaca, New York 14853, USA}
 \affiliation{Kavli Institute at Cornell for Nanoscale Science, Ithaca, New York 14853, USA}

\begin{abstract}
We report high-resolution angle-resolved photoemission studies of epitaxial thin films of the correlated $4d$ transition metal oxide ferromagnet SrRuO$_{3}$. The Fermi surface in the ferromagnetic state consists of well-defined Landau quasiparticles, exhibiting strong coupling to low-energy bosonic modes which contributes to the large effective masses observed by transport and thermodynamic measurements.  Upon warming the material through its Curie temperature, we observe a substantial decrease in quasiparticle coherence, but negligible changes in the ferromagnetic exchange splitting, suggesting that local moments play an important role in the ferromagnetism in SrRuO$_{3}$. 
\end{abstract}



\maketitle

The Ruddlesden-Popper series \cite{balz1955, ruddlesden1958} of ruthenates Sr$_{n+1}$Ru$_{n}$O$_{3n+1}$ displays remarkable electronic and magnetic properties where ferromagnetic tendencies are enhanced by increasing the number $n$ of RuO$_{2}$ sheets per unit cell. The single layer $n = 1$ compound Sr$_{2}$RuO$_{4}$ is proposed to be an exotic spin triplet superconductor with a time-reversal symmetry breaking ground state \cite{Mackenzie03}, while the $n = 2$ Sr$_{3}$Ru$_{2}$O$_{7}$ exhibits quantum critical metamagnetism \cite{Grigera2001} and an electronic nematic ground state near $B$ = 7.8 T. The series culminates in the pseudocubic, $n = \infty$ member, SrRuO$_{3}$, a correlated ferromagnet (FM) with Curie temperature ($T_{c}$) 160 K \cite{callaghan1966} and a low temperature moment of $1.4\ \mu_{B}$ \cite{longo1968}, rare for a $4d$ transition metal oxide. Because of its metallicity, magnetic properties, and epitaxial lattice match to other oxides, SrRuO$_{3}$ has become one of the central materials in oxide electronics \cite{Koster12} and has been utilized as a conductive electrode for ferroelectrics, Schottky diodes, magnetocalorics, and magnetoelectrics \cite{Junquera03, Fuji2005, Thota10, Niranjan09}. SrRuO$_{3}$ has even been proposed to support the existence of magnetic monopoles in $k$-space \cite{Fang03} or to form a building block for an oxide superlattice that supports a spin-polarized two-dimensional electron gas \cite{Alves12}. While its quasi-2D analogue Sr$_{2}$RuO$_{4}$ has been extensively studied using angle-resolved photoemission spectroscopy (ARPES), the inability to cleave single crystals of SrRuO$_{3}$, due to its pseudocubic structure, has meant that even a basic understanding of its low-energy electronic structure has remained out of reach. 

\begin{figure}[!hb]
	\includegraphics[width=1\columnwidth]{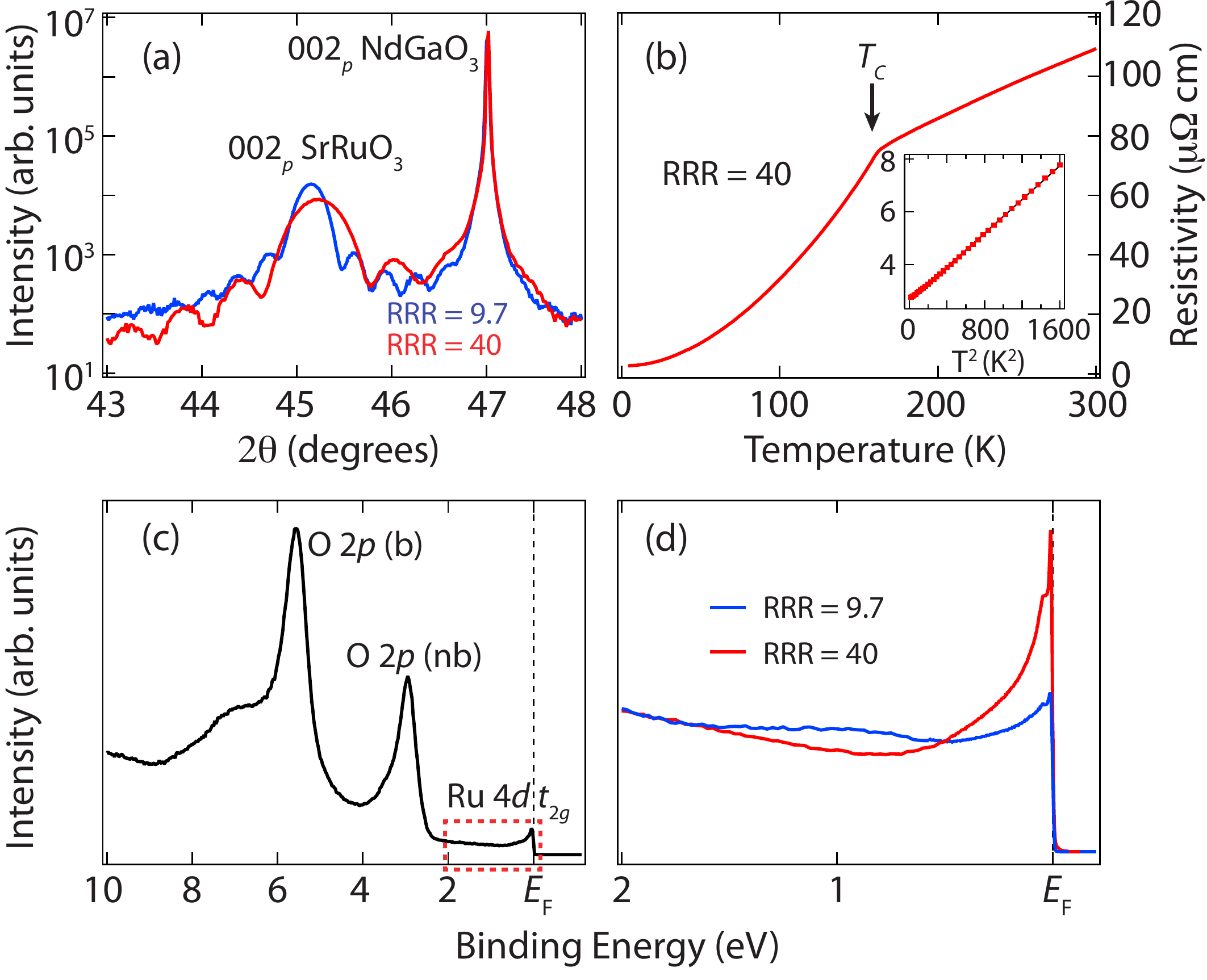}
	\caption{(a) X-ray diffraction $\theta-2\theta$ scan for two SrRuO$_{3}$ films showing an elongation of the $c$ lattice constant and lower RRR associated with Ru deficiency. The RRR=40 film is 18 nm thick and the RRR=9.7 film is 24 nm thick.   (b) Temperature-dependent resistivity for a SrRuO$_{3}$ film. (inset) Resistivity showing Fermi liquid $T^{2}$ dependence below 40 K. (c) Valence band photoemission near $(k_{x},k_{y})=0$ for SrRuO$_{3}$. (d) Comparison between the near-$E_{\rm F}$ spectra at $(0,0)$ from stoichiometric and Ru poor films. Spectra were normalized at 2 eV and were taken in the FM phase ($T =$ 20--30 K).}\label{ValenceBand}
\end{figure}

In this Letter, we report the first high-resolution ARPES measurements of SrRuO$_{3}$.  A mapping of the Fermi surface (FS) reveals a number of sheets composed of well-defined quasiparticle (QP) states.  We observe a prominent kink in the low-energy QP dispersion characteristic of coupling to bosonic modes indicating that the large effective masses reported in this material are a result of electron-boson interactions. Finally, we track the temperature dependence of the electronic structure from 20 K through $T_{c}$ and find that while the QPs lose coherence, they do not shift appreciably, indicating the ferromagnetic exchange splitting remains nonzero above $T_{c}$ in contrast to the expectations of itinerant Stoner FM.  These findings suggest that local moment FM is more appropriate for describing SrRuO$_{3}$.

Thin films of (001)$_{p}$ (where the subscript $p$ denotes pseudocubic indices) SrRuO$_{3}$ of thickness $\sim$20 nm were grown on (001)$_{p}$ NdGaO$_{3}$  single crystal substrates by molecular-beam epitaxy (MBE) in a dual-chamber Veeco GEN10 system. Films were grown in an oxidant (O$_{2}$ + 10\% O$_{3}$) background partial pressure of $8\times10^{-7}$  Torr at a substrate temperature of 800 $^{\circ}$C measured by pyrometer, and were monitored using reflection high-energy electron diffraction. After growth, samples were immediately ($<300$ seconds) transferred through ultrahigh vacuum to a high-resolution ARPES system consisting of a VG Scienta R4000 electron analyzer and a VUV5000 helium plasma discharge lamp and monochromator.  Measurements were performed using an energy resolution of 10 meV, He I$\alpha$ ($h\nu$ = 21.2 eV) photons, and base pressure of 8 $\times$ 10$^{-11}$ Torr. After performing ARPES, samples were characterized by low-energy electron diffraction, x-ray diffraction (XRD), and electrical transport \cite{supplemental}. 

As demonstrated by Siemons \emph{et al.} \cite{Siemons07}, the $c$-axis lattice constant is highly dependent on Ru stoichiometry. We found that only films with the correct Ru stoichiometry determined from XRD measurements of the $c$ axis [Fig. \ref{ValenceBand}(a)] would exhibit a high residual resistivity ratio (RRR = $\rho_{\rm 300K} / \rho_{\rm 4K}$) in transport measurements [Fig. \ref{ValenceBand}(b)] together with sharp, dispersive QP peaks near the Fermi level ($E_{\mathrm{F}}$) shown in Fig. \ref{ValenceBand}(d).  This strong dependence of spectral features on sample quality underscores the necessity of utilizing oxide MBE, which produces  thin films of SrRuO$_{3}$ with the highest RRR; the RRR of thin films grown by pulsed laser deposition or sputtering is typically 8 or smaller \cite{kacedon1997,chu1996}.  In Fig. \ref{ValenceBand}(c), the valence band of SrRuO$_{3}$ is shown. By comparison with existing density functional calculations \cite{fujioka1997}, we can identify the features between 3 and 7 eV as primarily O $2p$ bonding (b) and nonbonding (nb) states, while the peak near $E_{\rm F}$ can be assigned to the Ru $4d$ $t_{2g}$ orbitals, consistent with results from angle-integrated photoemission \cite{Kim05}.

\begin{figure}[!t]
	\includegraphics[width=1\columnwidth]{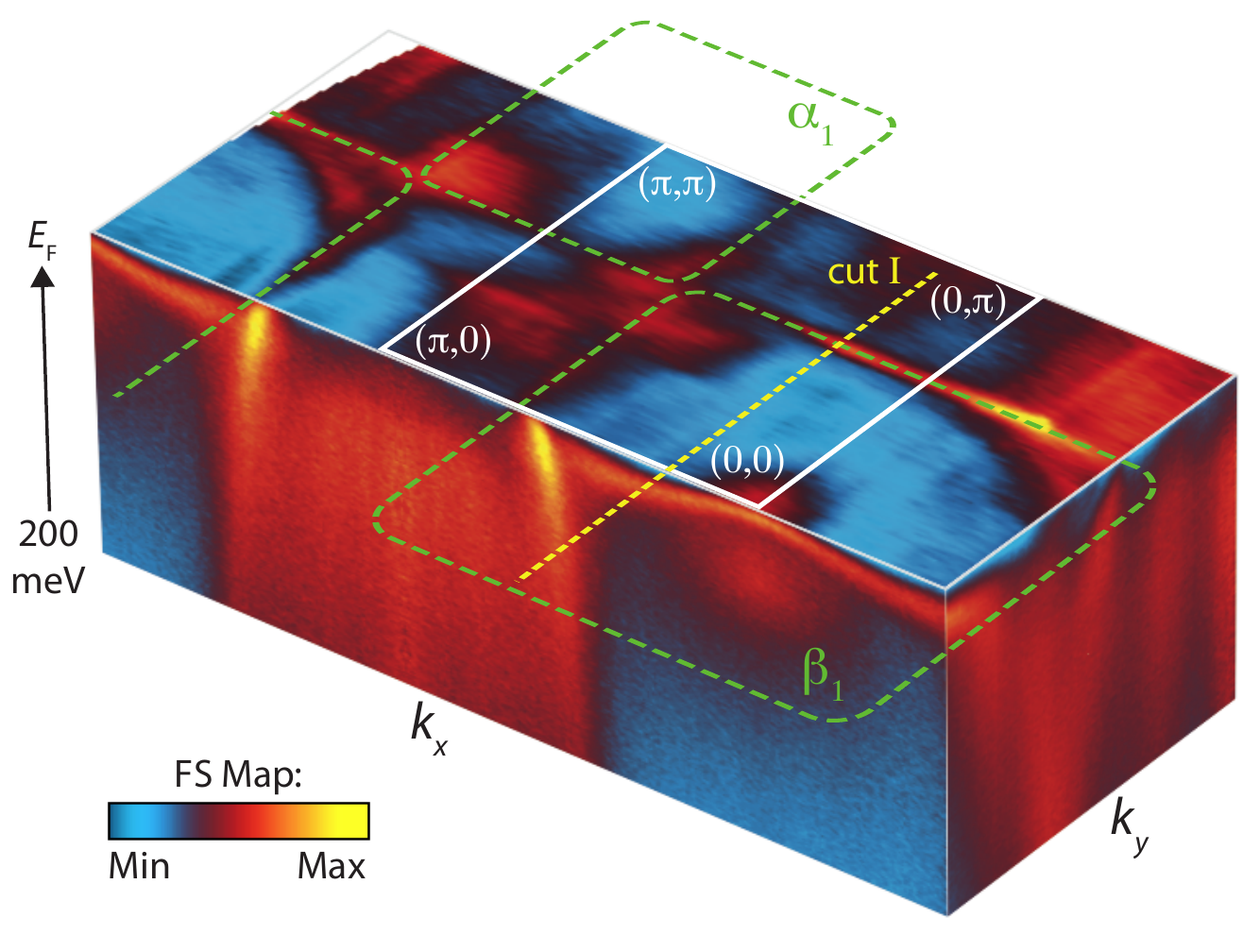}
	\caption{Unsymmetrized FS map for SrRuO$_{3}$ at $T = 20$ K integrated within $E_{\mathrm{F}}\ \pm$ 5 meV, along with $E$ vs. $k_{x}$, $k_{y}$ spectra illustrating the underlying band structure.    The FS was normalized to a linearly increasing background in $k_{x}$ to account for a slowly varying change in net intensity.  $E$ vs $k$ data were taken at symmetry equivalent locations in the Brillouin zone, and are symmetrized for values of $k_{x}>\pi/a$ and plotted on separate intensity scales (not shown) for illustrative purposes.
	\label{fig:FSMaps}}
\end{figure}

\begin{figure*}[]
	\includegraphics[width=1\linewidth]{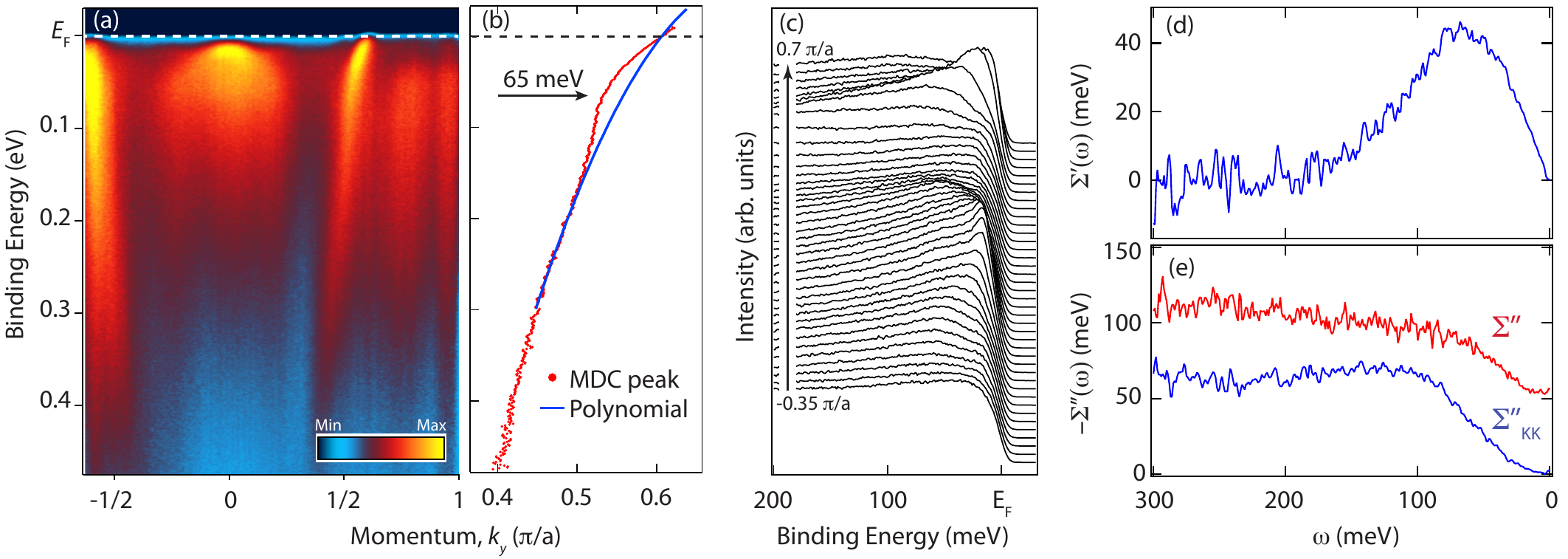}
	\caption{(a) ARPES spectral intensity plot taken at $T=20$ K along cut I (see Fig. \ref{fig:FSMaps}). (b) Extracted MDC dispersion for the $\beta_{1}$ band along with a quadratic polynomial fit approximating the bare band dispersion. (c) Energy distribution curves showing QP peaks in SrRuO$_{3}$. (d). Real part of the self-energy $\Sigma^{\prime}(\omega)$ extracted from the MDC dispersion and polynomial fit in (b).  (e)  Imaginary part of the self-energy $\Sigma^{\prime\prime}(\omega)$ computed from MDC widths, along with the Kramers-Kronig transformation of $\Sigma^{\prime}(\omega)$ denoted  $\Sigma^{\prime\prime}_{\rm KK}(\omega)$.}\label{fig:kink}
\end{figure*}

The fermiology of FM SrRuO$_{3}$ is expected to be complex. In Sr$_{2}$RuO$_{4}$, each Ru atom contributes three partially filled $4d$ $t_{2g}$ bands. Because of octahedral rotations, there are four Ru atoms for each SrRuO$_{3}$ unit cell, resulting in 12 $t_{2g}$ orbitals, which would be doubled to 24 from the FM splitting. Additionally, calculations by Santi and Jarlborg \cite{santi1997} suggest that the $e_{g}$ orbitals may also be partially filled. In Fig. \ref{fig:FSMaps} we show a Fermi surface intensity map along with the underlying $E$ vs $k$ spectra. A number of FS sheets is evident, including two large rectangular sheets which are reminiscent of the quasi-1D, $d_{xz,yz}$-derived $\alpha$ and $\beta$ sheets present in Sr$_{2}$RuO$_{4}$, which we denote as $\alpha_{1}$ and $\beta_{1}$. Particularly around $(2\pi/3, 2\pi/3)$ (with respect to the pseudocubic Brillouin zone), near where the $\alpha_{1}$ and $\beta_{1}$ sheets meet, we observe a large number of band crossings, which result in numerous small FS pockets. Earlier Shubnikov--de Haas (SdH) \cite{Mackenzie98} and de Haas--van Alphen (dHvA) \cite{Alex05} measurements both resolved a coexistence of large and very small FS sheets, which are consistent with the large $\beta_{1}$ sheet and the small pockets that we observe near $(2\pi/3, 2\pi/3)$.  We find the area enclosed by the $\beta_{1}$ sheet to be 1.02 \AA$^{-2}$, in good agreement with the 10.5 kT (1.00  \AA$^{-2}$) orbit reported in Ref. \onlinecite{Alex05}.  The area enclosed by the $\alpha_{1}$ pocket is approximately 0.37 \AA$^{-2}$, close to the 3.5 kT (0.33 \AA$^{-2}$) orbit in Ref. \onlinecite{Mackenzie98}.  While our measurements probe only a single value of the out-of-plane momentum ($k_{z}$), we expect them to accurately represent the features of the $\alpha_{1}$ and $\beta_{1}$ sheets, which likely exhibit little $k_{z}$ dispersion along large sections of the Brillouin zone due to their apparently quasi-1D nature.  A detailed discussion of the complete, complex three-dimensional fermiology will follow in a future paper. 

In order to address the nature of the many-body interactions in SrRuO$_{3}$, we focus on the momentum region marked as cut I in Fig. \ref{fig:FSMaps}, as this is where the $\beta_{1}$ band appears to be most separated from the other bands, and thus is best suited for a detailed analysis. The spectra in this region are shown in Figs. \ref{fig:kink}(a) and \ref{fig:kink}(c) and exhibit a sharp QP peak at $E_{\rm F}$, deep in the $T^{2}$ Fermi liquid (FL) regime [Fig. \ref{ValenceBand}(b)]. This is also consistent with SdH measurements \cite{Mackenzie98} that indicated the low-temperature ground state of SrRuO$_{3}$ is indeed FL-like. We extract the dispersion of the $\beta_{1}$ band using a momentum distribution curve (MDC) analysis shown in Fig. \ref{fig:kink}(b). The dispersion is then fit to a quadratic polynomial to account for the intrinsic curvature of the band over a wide energy range. An abrupt kink in the dispersion is evident around 65-meV binding energy. We also observe a prominent increase in the MDC width ($W_{k}$) at the same energy, indicating a sudden increase in the imaginary part of the self-energy $\Sigma^{\prime\prime}(\omega)=-v_{\rm F} W_{k}$ shown in Fig. \ref{fig:kink}(e).  The observation of a sharp kink in the dispersion with a corresponding increase in the scattering rate is a classic signature of strong electron-boson interactions.  To further support this assignment, we extract the real part of the self-energy $\Sigma'(\omega)$ by subtracting the smooth polynomial dispersion in Fig. \ref{fig:kink}(b), and perform a Kramers-Kronig transformation (KKT) of $\Sigma'(\omega)$ to obtain $\Sigma''(\omega)$ \cite{KKT}. We show in Figs. \ref{fig:kink}(d) and \ref{fig:kink}(e) that the self-energy extracted directly from the dispersion $\Sigma^{\prime\prime}(\omega)$ matches closely to the self-energy computed from the KKT $\Sigma_{\rm KK}^{\prime\prime}(\omega)$, aside from a constant offset and a slight decrease in the change in scattering rate with $\omega$, which we attribute to impurity scattering and finite instrumental resolution effects, respectively. This close agreement indicates that our self-energy analysis is not only self-consistent, but that this kink arises from many-body interactions and is not an artifact of band crossings or hybridization.

Electronic specific heat measurements \cite{Allen96} show a large mass renormalization of $m^{\ast} / m_{b} \approx 4$, where the band mass $m_{b}$ was determined from density functional calculations. Although early angle-integrated photoemission \cite{fujioka1997} and optical conductivity measurements \cite{Ahn99} suggested that electron-electron interactions might be responsible for the large effective masses, later optical \cite{Lee01} and angle-integrated photoemission measurements \cite{Maiti05} indicate that electron-electron interactions could be relatively weak. Our observation of an abrupt kink in the $\beta_{1}$ band provides direct evidence that strong electron-boson coupling in SrRuO$_{3}$ is a dominant factor in the large observed effective masses. This is consistent with recent first principles calculations which suggest weak electronic correlations in SrRuO$_{3}$ \cite{Etz12}. By comparing the bare velocity $v_{F}$ (from our polynomial fit) to the renormalized $v_{F}^{\ast}$ (computed from the MDC peak dispersion), we obtain a ratio of $v_{F} / v_{F}^{\ast} = 1.9 \pm 0.2$; by performing a more conventional analysis where we fit straight lines to the dispersion between 100 and 150 meV and between $\pm$10 meV of $E_{\rm F}$, we extract an even larger quantity of $v_{HE} / v_{F}^{\ast} = 5.5 \pm 0.5$. We believe that the former analysis provides a more accurate estimate of the true velocity renormalization and in a weak-coupling scenario results in a mass renormalization, $m^{\ast} / m_{b}$, of 1.9, which places SrRuO$_{3}$ well into the strong-coupling regime. Averaging this value around the $\beta_{1}$ sheet, we deduce a sheet-averaged effective mass $m^{\ast}=\hbar^{2}\langle k_{\rm F}\rangle/v_{F}^{\ast}\approx3.9\pm0.7\ m_{e}$, in excellent agreement with the dHvA value of $4.1\pm0.1\ m_{e}$ \cite{Alex05} for this sheet.

\begin{figure}
	\includegraphics[width=1\columnwidth]{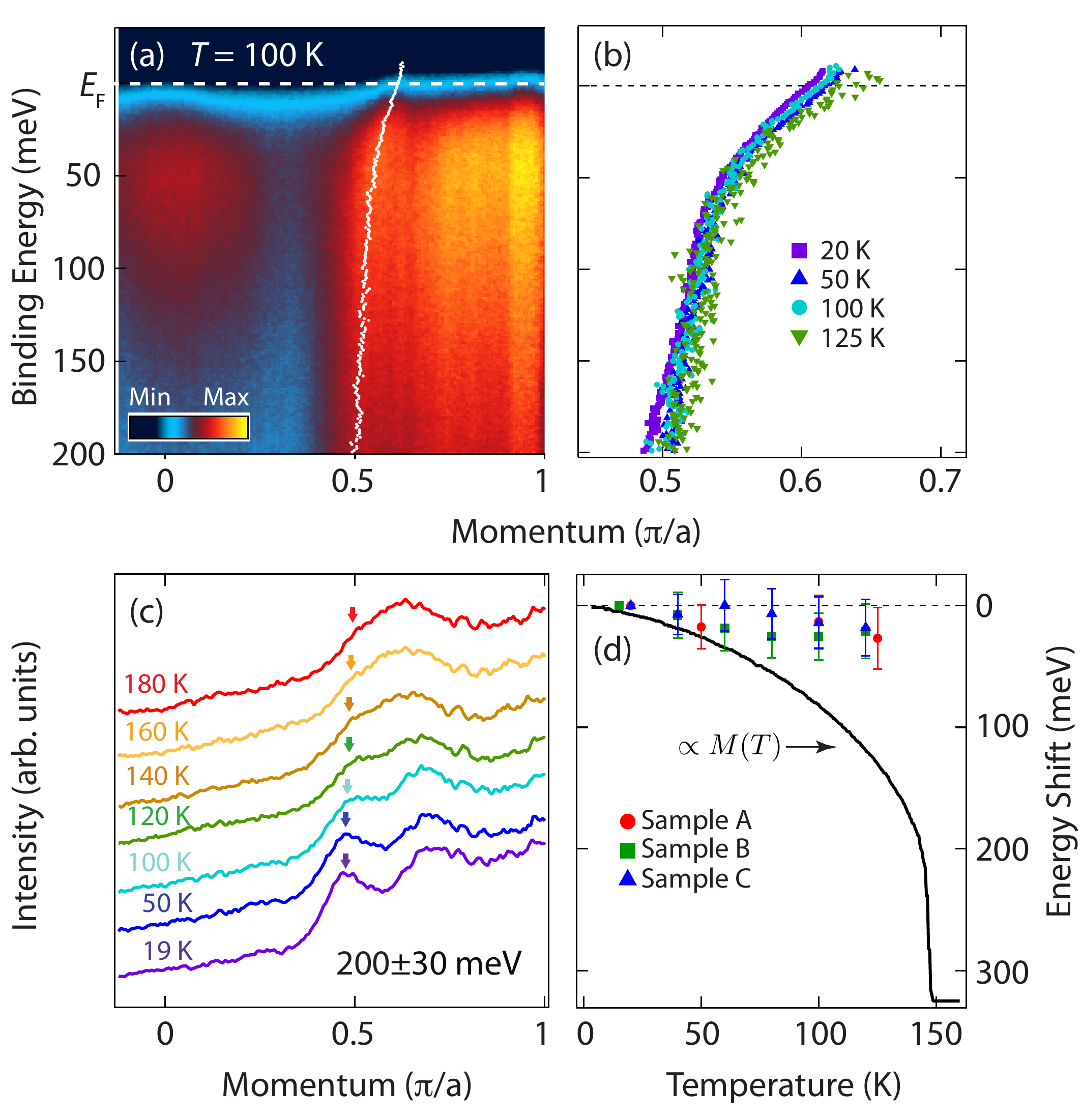}
	\caption{\label{fig:temperature} (a) ARPES spectral intensity plot taken at $T =$ 100 K along cut I in Fig. \ref{fig:FSMaps}. Overlaid is the extracted MDC dispersion.  (b) MDC extracted dispersion of the $\beta_{1}$ band at temperatures between 20 and 125 K.  (c) MDCs (offset for clarity) showing the evolution of the $\beta_{1}$ band through $T_{c}$.  (d) Temperature-dependent $\beta_{1}$ band energy shift from multiple ARPES samples and expected energy shift from Stoner FM.   Magnetization data were taken from Ref. \onlinecite{Klein96}.}
\end{figure}

We now discuss the temperature dependence of the electronic structure, concentrating again on the $\beta_{1}$ band.  As the temperature is increased, we observe that the sharp QP features on the $\beta_{1}$ sheet lose coherence and become significantly broadened, as shown in Fig. \ref{fig:temperature}(a).  Nevertheless, we can still reliably track this band using MDCs from low temperature (19 K) up towards $T_{c}$.  In Fig. \ref{fig:temperature}(c) we present MDCs at 200 meV binding energy which clearly show the $\beta_{1}$ band at all temperatures.  Even above $T_{c}$ this band is resolvable as a shoulder to the feature located at $0.7\ \pi/a$.  We observe the same behavior both upon warming and cooling the sample, excluding the possibility of surface aging as being responsible for this effect.  Up to temperatures of 125 K we can reliably track the band dispersion from 200 meV all the way to $E_{\rm F}$, as shown in Fig. \ref{fig:temperature}(b). We observe negligible changes in the shape of dispersion, and the kink persists throughout the entire temperature range.

By comparing the average shift in momentum $\Delta k$ for the extracted dispersion in Fig \ref{fig:temperature}(b), we can compute the temperature-dependent energy shift of the $\beta_{1}$ band $\Delta E = \Delta k\ dE/dk$, presented in Fig. \ref{fig:temperature}(d).  We find that at 125 K, the $\beta_{1}$ band has shifted by only $27\pm25$ meV from its low temperature value.  Within the context of the Stoner model  \cite{Stoner1938}, the exchange splitting of an itinerant FM should decrease from its saturation value at low temperatures to zero at $T_{c}$, varying proportionally to the bulk magnetization.  For comparison, we also include in Fig. \ref{fig:temperature}(d) the temperature-dependent magnetization \cite{Klein96} which has been scaled to half of the density functional theory-predicted exchange splitting (325 meV, the expected shift per spin population) \cite{Singh96}, showing the strong deviation of the bands in SrRuO$_{3}$ from the expectations of a simple Stoner model. We also did not observe substantial shifts in other bands, although we could not extract their dispersions as reliably as the $\beta_{1}$ band. These results are consistent with earlier bulk-sensitive optical measurements where a corresponding small shift of only 40 meV (of which only a maximum of 10 meV could be attributed to the exchange splitting) was reported upon warming to 140 K \cite{Dodge99PRB}, and with recent magnetization measurements which showed deviations from the Stoner model across a series of perovskite ruthenates, $A$RuO$_{3}$ ($A$ = Ca, Sr, Ba) \cite{Cheng2012}. This agreement between the optical, magnetic, and ARPES measurements clearly points towards a scenario where the magnetism has a local character.

Previous magnetic measurements \cite{Fukunaga94} indicate a ratio of the Curie-Weiss moment to the saturation moment of 1.3, placing SrRuO$_{3}$ in the category of a more localized moment FM, similar to Fe or Ni. The existing experimental literature on Ni is somewhat complex and possibly conflicting, with some reports showing little change in the spin polarization of Ni through $T_{c}$ \cite{Sinkovic97}, while other work appears to show an exchange splitting which decreases to zero at $T_{c}$ \cite{Kreutz98}. Spin-polarized photoemission on Fe reveals exchange split bands with little temperature dependence when heated through $T_{c}$ \cite{Kisker84}. Such observations have led to propositions that these metals retain their local moment above $T_{c}$, but lose long-range magnetic order resulting from transverse spin fluctuations \cite{BaberschkeBook}, which could be consistent with our ARPES results for SrRuO$_{3}$. 

In summary, our ARPES measurements provide the first insights into the momentum-resolved electronic structure of SrRuO$_{3}$. Our observation of a kink in the quasiparticle dispersion conclusively demonstrates that strong electron-boson interactions have an important role in the large mass renormalization in SrRuO$_{3}$. Furthermore, our temperature-dependent measurements of the electronic structure suggest that a simple Stoner model cannot adequately describe the magnetism in SrRuO$_{3}$ and that local moments may play an important role.  This new understanding of the electronic structure and quasiparticle interactions is an important step in realizing novel oxide interfaces, electronics, and devices based on SrRuO$_{3}$. 

We would like to thank T. Heeg for assistance with the thin film growth, D.J. Singh for sharing unpublished band structure calculations, and T.W. Noh, S. Chatterjee, P.D.C. King, and M. Uchida for helpful discussions. This work was supported by the National Science Foundation through a CAREER Grant No. DMR-0847385 and the Materials Research Science and Engineering Centers (MRSEC) program (Grant No. DMR-1120296, Cornell Center for Materials Research), as well as a Research Corporation Cottrell Scholars Grant No. 20025. D.E.S. acknowledges support from the National Science Foundation under Grant No. DGE-0707428 and NSF IGERT under Grant No. DGE-0654193.  C.A. and D.G.S. acknowledge support from AFOSR Grant No. FA9550-10-1-0524.  E.J.M. acknowledges NSERC for PGS support.

\end{document}